\documentclass[traditabstract]{aa}

\usepackage{graphicx,hyperref}
\usepackage{rotating,lscape}
\newcommand{\teff}{T_{\text{eff}}}
\usepackage{txfonts}
\begin{document} 
\title{The elderly among the oldest: new evidence for extremely metal-poor RR Lyrae stars\thanks{Based on observations acquired at the Telescopio Nazionale Galileo under program A43DDT3, on a DDT program with PEPSI at LBT (2021-2022, PI Crestani) and on VLT ESO programs 69.C-0423(A) and 165.N-0276(A).}}
\author{V. D'Orazi
          \inst{1,2}
          \and
            V. Braga
          \inst{3}
            \and
            G. Bono
          \inst{1,3}
          \and
        M. Fabrizio
          \inst{3}
           \and
        G. Fiorentino
          \inst{3}
          \and
            N. Storm
           \inst{4}
         \and
          A. Pietrinferni
          \inst{5}
          \and
          C. Sneden
          \inst{6}
          \and
          M. S\'anchez-Benavente
          \inst{7,8}
          \and
          M. Monelli
          \inst{7,8,3}
          \and
          F. Sestito  
          \inst{9}
          \and
          H. J\"{o}nsson
          \inst{10}
          \and
          S. Buder
          \inst{11,12}
        \and
         A. Bobrick
         \inst{13}
        \and
         G. Iorio
         \inst{14}
          \and
            N. Matsunaga
          \inst{15,16}
          \and
            M. Marconi 
            \inst{17}
          \and
           M. Marengo
           \inst{18}
           \and
           C. E. Mart\'inez-V\'azquez
           \inst{19}
           \and
           J. Mullen
           \inst{20}
           \and
           M. Takayama 
           \inst{21}
           \and 
           V. Testa
           \inst{3}
           \and
           F. Cusano
           \inst{22}
           \and
           J. Crestani
           \inst{1,3}
          }

   \institute{Department of Physics, University of Rome Tor Vergata, 
   via della ricerca scientifica 1, 00133, Rome, Italy\\
              \email{vdorazi@roma2.infn.it}
              \and
             INAF -- Osservatorio Astronomico di Padova, vicolo dell'Osservatorio 5,
             35122, Padova, Italy
             \and
             INAF -- Osservatorio Astronomico di Roma, via Frascati 33, Monte porzio catone, Italy
            \and
             Max Planck Institute for Astronomy,  K\"{o}nigstuhl 17, 69117 Heidelberg, Germany
             \and
             INAF -- Osservatorio Astronomico d'Abruzzo, via Maggini sn, I-64100 Teramo, Italy 
             \and
             Department of Astronomy and McDonald Observatory, The University of Texas, Austin, TX 78712, USA
             \and
             IAC - Instituto de Astrofisica de Canaria, calle Via Lactea s/n, E-38205 La Laguna, Tenerife, Spain 
             \and
             Departamento de Astrofısica, Universidad de La Laguna, E-38206 La La guna, Tenerife , Spain 
            \and
             Centre for Astrophysics Research, Department of Physics, Astronomy and Mathematics, University of Hertfordshire, Hatfield, AL10 9AB, UK.
             \and
             Materials Science and Applied Mathematics, Malm\"o University, SE-205 06 Malm\"{o}, Sweden
             \and
             Research School of Astronomy \& Astrophysics, Australian National University, ACT 2611, Australia
             \and
            Center of Excellence for Astrophysics in Three Dimensions (ASTRO-3D), Australia
             \and
             Physics Department, Technion –Israel Institute of Technology, Haifa 32000, Israel 
             \and
             Departament de F\'{i}sica Qua\'antica i Astrof\'{i}sica, Institut de Cie\'encies del Cosmos, Universitat de Barcelona, Marti\'i i Franque\'es 1, 08028 Barcelona, Spain
              \and
             Department of Astronomy, School of Science,The University of Tokyo, 7-3-1 Hongo, Bunkyo-ku, Tokyo 113-0033, Japan
             \and
             Laboratory of Infrared High-resolution spectroscopy (LiH), Koyama Astronomical Observatory, Kyoto Sangyo University, Motoyama, Kamigamo, Kita-ku, Kyoto 603-8555, Japan
             \and
             INAF -- Osservatorio Astronomico di Capodimonte, salita Moiariello 16, I-80131 Napoli, Italy 
             \and
             Department of Physics, Florida State University, 77 Chieftain Way, Tallahassee, FL32306, USA 
             \and
             International Gemini Observatory/NSF NOIRLab, 670 N. A'ohoku Place, Hilo, Hawai'i, 96720, USA
             \and
            Department of Physics and Astronomy, Vanderbilt University, Nashville, TN 37240, USA
             \and
             Nishi-Harima Astronomical Observatory, Center for Astronomy,
            University of Hyogo, 407-2, Nishigaichi, Sayo-cho, Hyogo 679-5313, Japan
             \and
             INAF -- Osservatorio Astrofisica e Scienza dello Spazio, via Gobetti 93/3, Bologna, Italy 
                }

   \date{Received: 28 November 2024 | Accepted  6 January 2025}

 
  \abstract 
  {We performed a detailed spectroscopic analysis of three extremely metal-poor RR Lyrae stars, exploring uncharted territories at these low metallicities for this class of stars. Using high-resolution spectra acquired with HARPS-N at TNG, UVES at VLT, and PEPSI at LBT, and employing Non-Local Thermodynamic Equilibrium (NLTE) spectral synthesis calculations, we provide abundance measurements for Fe, Al, Mg, Ca, Ti, Mn, and Sr.
  Our findings indicate that the stars have metallicities of [Fe/H] = $-3.40 \pm 0.05$, $-3.28 \pm 0.02$, and $-2.77 \pm 0.05$ for HD 331986, DO Hya, and BPS CS 30317-056, respectively. Additionally, we derived their kinematic and dynamical properties to gain insights into their origins. Interestingly, the kinematics of one star (HD 331986) is consistent with the Galactic disc, while the others exhibit Galactic halo kinematics, albeit with distinct chemical signatures. We compared the [Al/Fe] and [Mg/Mn] ratios of the current targets with recent literature estimates to determine whether these stars were either accreted or formed in situ, finding that the adopted chemical diagnostics are ineffective at low metallicities  ([Fe/H] $\lesssim -$1.5). Finally, the established horizontal branch evolutionary models, indicating that these stars arrive at hotter temperatures on the Zero-Age Horizontal Branch (ZAHB) and then transition into RR Lyrae stars as they evolve, fully support the existence of such low-metallicity RR Lyrae stars. As a consequence, we can anticipate detecting more of them when larger samples of spectra become available from upcoming extensive observational campaigns.}


    \maketitle

\section{Introduction}

    Since the publication of the seminal paper by \cite{baade1958}, RR Lyrae stars (RRLs) have become reliable indicators of ancient stellar populations \citep[and references therein]{layden1996, sneden2017, preston2019, braga2021}.
    These stars are radial variables that pulsate in various modes: the fundamental mode (RRab), the first overtone mode (RRc), and both the fundamental and first overtone modes simultaneously (RRd). Their periods typically range from a few hours to just under a day. RRLs are low-mass horizontal-branch (HB) stars, with masses between $\approx$ 0.50 and 0.85 M$_\odot$ \citep{marconi2015,marsakov2019,bobrick2024}, which burn helium in their cores and hydrogen in a surrounding shell. Compared to similar stellar tracers, RRLs offer three key advantages. 
{\em i)} They are ubiquitous, meaning they have been identified in 
all stellar systems hosting old stellar populations. 
{\em ii)} They are accurate primary distance indicators and obey the well-defined Period-Luminosity (PL) relation for wavelengths longer than the $R$-band. 
{\em iii)} They can be easily identified due to the distinctive coupling between the shape of their light curves and the pulsation period. There are some classification difficulties 
for RRc variables that overlap with geometrical (binaries) and Delta Scuti variables. Still, pulsation observable (the ratio of luminosity amplitudes) can be adopted to improve their identification \citep{mullen2022}. The main drawback is that they are several magnitudes fainter than Miras, Classical Cepheids, and type {\sc ii} Cepheids. 

Baade was the first to observe differences in the light curves of RRLs between those located in the bulge and those in globular clusters and the halo. Even before obtaining spectroscopic measurements, he proposed that RRLs in the bulge were more metal-rich. This hypothesis was later confirmed by the pioneering spectroscopic studies conducted by \citet{preston1959,preston1964}, which demonstrated that many field RRLs were indeed metal-rich and exhibited disc kinematics.
The kinematics were more typical of thick disc stars and it took over thirty years until another critical spectroscopic investigation by \citet{layden1995} identified, through the $\Delta$S method\footnote{This method exploits the strengths of Ca {\sc ii h,k} and hydrogen Balmer lines to estimate the metallicity (see e.g. \citealt{crestani2021b}, and references therein).}, metal-rich RRLs with thin disc kinematics. A few years earlier, \citet{walkerterndrup1991} utilised the $\Delta$S method to provide the first evidence of bulge RRLs approaching solar iron abundance. Subsequently, \cite{wallerstein2011} employed the Ca {\sc ii} triplet and confirmed the existence of solar-metallicity RRLs.

 Recent spectroscopic studies have shown that RRLs exhibit $\alpha$-enhanced abundance patterns in metal-poor stars ([Fe/H] $\leq$ -1) and display a composition closer to the solar mixture in more metal-rich stars \citep{for2011, chadid2017, sneden2017, magurno2018}. A similar conclusion was also reached by \citet[][and references therein]{marsakov2019}. Recent studies by \citet{crestani2021a,crestani2021b,dorazi2024} have contributed to this field by providing accurate abundances for several $\alpha$ elements, such as magnesium, calcium, and titanium. Using relatively large(r) samples, these investigations revealed that RRLs with iron abundances close to solar levels display sub-solar $\alpha$-element abundances.

The results from these studies indicate that RRLs span over 2.5 dex in iron abundance and exhibit a wide range in $\alpha$-element abundances. In this context, the low-end metallicity distribution is mainly represented by RRLs in Galactic globular clusters, particularly those in M68, M92, and M15, with iron abundances around $-$2.5 dex. Unexpected discoveries of extremely metal-poor (EMP, [Fe/H] $\lesssim -$2.7) RRLs were made by \citet{hansen2011}, who identified two field RRLs with iron abundances of $\approx -$2.8, and by \citet{crestani2021b}, who determined that DO~Hya also shares a similar iron abundance. Surprisingly, \citet{matsunaga2022} discovered a field RRL with disc kinematics and an iron abundance possibly even more metal-poor than [Fe/H]$\approx -$3. The lack of high-resolution optical spectra initially limited abundance analyses of these stars. 
This study aims to explore the lower limits of metallicity for field RRLs, a poorly explored region compared to our understanding of RR Lyrae in clusters.  We determine detailed elemental abundances for these stars and, through dynamical and chemical analyses, investigate their origin and association with specific Galactic populations. This paper presents the first detailed spectroscopic analysis of EMP field RRLs, which exist in a metallicity range even lower than that typically found in globular clusters.

\section{Spectroscopic analysis}

Our observational efforts focused on the three currently known EMP RRLs field stars: HD 331986, DO Hya, and BPS CS 30317-056. We employed the HARPS-N at the Telescopio Nazionale Galileo (TNG, Roque de los Muchachos, Spain) spectroscopic data of HD 331986 (also known as Matsunaga's star; \citealt{matsunaga2022}) that exhibit a nominal resolution of R=115,000 \citep{cosentino2014S}. This analysis focused on observations conducted near the pulsation phase $\phi$ = 0.87, which occurs at the onset of rising light \footnote{Note that we adopted, as reference epoch, the time of the mean magnitude on the rising branch and not the typical time of maximum light \citep{inno15,braga2021}}. At this stage, effective temperatures are still close to their minimum values within the pulsation cycle \citep{magurno2019}, enabling us to enhance the detectability of spectral features. The selected spectrum, recorded on 24 July 2021, offers a spectral range from 3900 to 6900 $\AA$ and a signal-to-noise ratio (SNR) of 36 per pixel around 6000 \AA.~ The $\teff$ was determined using a Non-Local Thermodynamic Equilibrium (NLTE) approach for H$_{\alpha}$ line fitting, and the surface gravity was inferred through the ionisation equilibrium of iron absorption lines. Our analysis used the Python wrapper for {\tt TurboSpectrum 2020}, TSFitPy \citep{gerber2023, storm2023}, which enables the calculation of NLTE synthetic spectra in real-time and accounts for best-fit determinations against observed spectra using the Nelder-Mead algorithm (e.g. \citealt{storm2023}). The microturbulent velocity ($V_{\rm{mic}}$) was determined by removing any evident trends between the Reduced Equivalent Width (REW) and the NLTE iron abundances derived from Fe I lines. All parameters and abundances reported in this study are based on NLTE calculations; our recent publication \citep{dorazi2024} provides a comprehensive discussion of these methodologies and their implications. For completeness, we also reviewed additional spectra obtained with HARPS-N (three spectra at hotter phases of $\phi$=0.07, 0.431, 0.636) and PEPSI at the Large Binocular Telescope \citep{2015AN....336..324S}, which consistently corroborated our metallicity estimates within an uncertainty margin of 0.1 dex. 

\begin{figure*}
    \centering
    \includegraphics[width=0.95\textwidth]{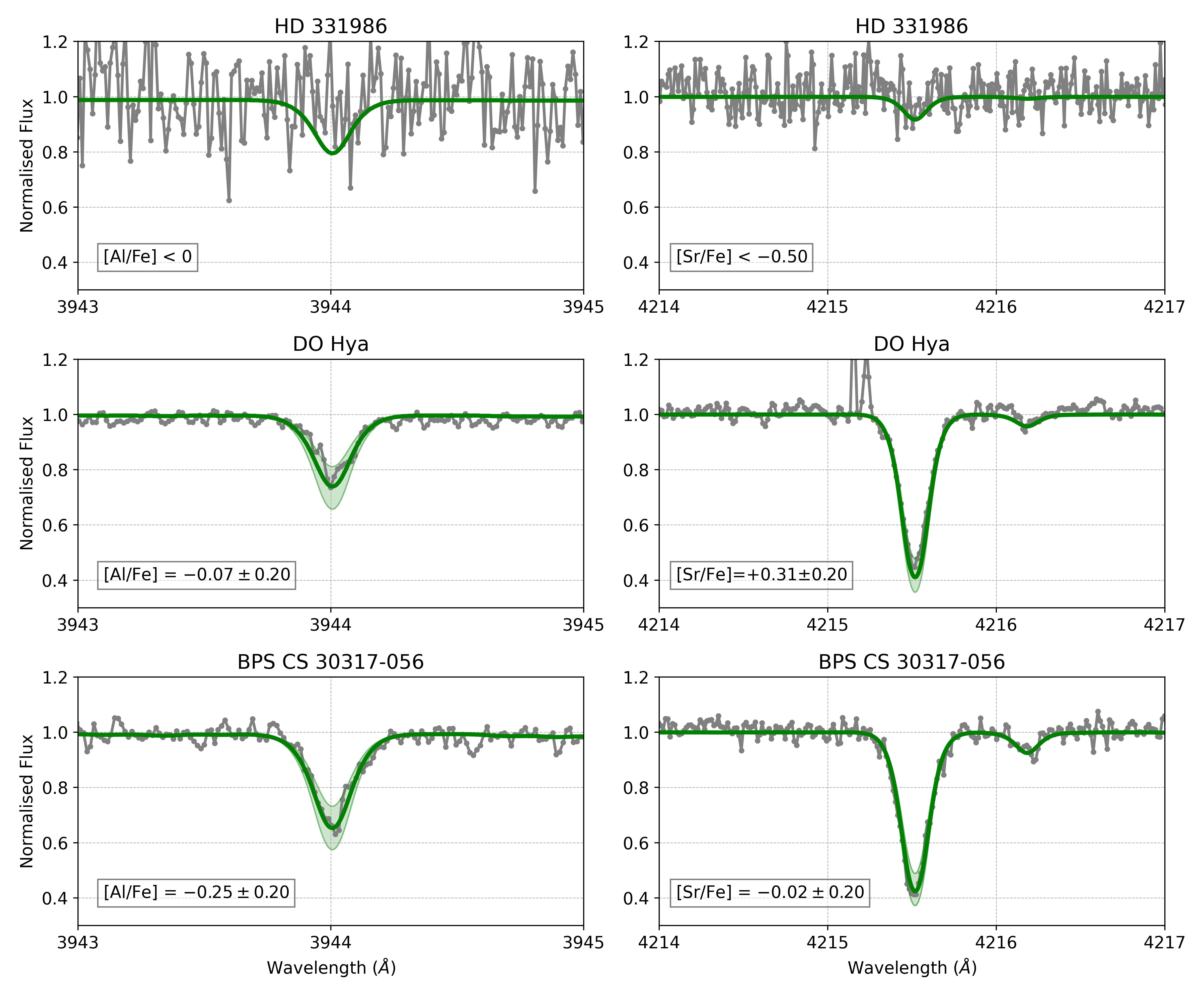}
    \caption{Comparison of the observed (grey) and synthetic (green) spectra. For HD 331986 (upper panel), we present the upper limit synthesis, while for DO Hya (middle panel) and BPS CS 30317-056 (lower panel), we display the best-fit results along with a variation of $\pm 0.2\,\mathrm{dex}$ (shaded areas).}
    \label{fig:synth}
\end{figure*}

To mitigate systematic errors and standardise the metallicity scale for the two most metal-deficient RRLs ever reported, we re-evaluated the spectrum of DO Hya, initially included by \cite{crestani2021a}; observational details regarding this spectrum are provided in the referenced study. Our elemental analysis extended beyond iron, encompassing determinations of abundances for aluminium (Al), magnesium (Mg), calcium (Ca), titanium (Ti), manganese (Mn), and strontium (Sr), with NLTE spectral fitting for all the species under scrutiny (see details provided in Table~\ref{table:abundances}).
An example of the spectral synthesis technique for the 
Al~{\sc i} line at 3944~\AA~  and the Sr~{\sc ii} line at 4077 \AA\ is shown in Figure~\ref{fig:synth}. Unfortunately, we were unable to determine the carbon abundances for our sample stars. Nevertheless, our spectral analysis suggests that substantial enhancements in the [C/Fe] ratios (like those inferred in Carbon-Enhanced Metal-Poor stars -CEMP) are unlikely, as we establish upper limits of [C/Fe] $\lesssim$ 1. For completeness, we report that \cite{matsunaga2022} established an upper limit of [C/H] $\leq$ -2.5, which aligns well with our estimate when using our metallicity value of [Fe/H] = $-3.40$. It is worth recalling that in their study they could only determine an upper limit on iron abundance of [Fe/H] $\leq -$2.5. These findings indicate a distinct behaviour of our three EMP RRLs compared to the CEMP RRLs identified by \cite{kennedy2014}. Notably, they inferred [C/Fe] = +1.35 for WY Vir, which has a metallicity of [Fe/H] = $-2.65$.

In the Introduction, we referred to the work by \cite{hansen2011}, who identified two exceptionally metal-poor RRLs: BPS CS 30317-056 and BPS CS 22881-039. The re-evaluation focused on BPS CS 30317-056 and is discussed further in the subsequent sections. Our findings align with those initially reported, considering observational uncertainties. 
Conversely, we excluded BPS CS-22881-039 from our analysis due to its significantly broad spectral features. This broadening, likely due to high macroturbulence velocity, introduces large uncertainties in metallicity measurements, potentially exceeding 0.3 dex. This assessment aligns with findings from \cite{crestani2021a}, who, through analysis of two spectra at different phases, reported metallicity values of [Fe/H] = $-$2.33 and [Fe/H] = $-$2.66. In adherence to a conservative approach that prioritises the inclusion of only robust abundance data, we decided to omit BPS CS-22881-039 from our discussion.


\section{Results and discussion}
This paper presents the first comprehensive chemical and dynamical analysis of the three most metal-poor  RRLs ever detected.  Our study examines the stars' origin (as detailed in the following Section \ref{sec: chemodynamics}) and compares observational data with theoretical evolutionary models to better understand their fundamental properties (see Section \ref{sec:evolutionary}).

\subsection{Chemodynamics of our EMP RR Lyrae stars}\label{sec: chemodynamics}

We have determined orbital parameters using the {\tt galpy} code \citep{Bovy2015}, adopting the \citet{mcmillan17} Galactic potential. For these calculations, we adopted the Sun's Galactocentric location as $R_\sun = 8.122$~kpc \citep{gravitycollaboration18} and $z_\sun=20.8$~pc \citep{bennett19}, while as 3D components of the Sun's velocity, we used $(U_\sun,V_\sun,W_\sun)=(12.9,245.6,7.78)$~km~s$^{-1}$ \citep{drimmelandpoggio18}. The circularity of the orbits $\lambda_z$ \citep[see e.g.][]{massari2019}, defined as the angular momentum along the $z$-axis ($J_z$) normalised by the angular momentum of a circular orbit with the same binding energy ($E$), was used to identify the characteristics of the orbits \citep[for details see][]{dorazi2024}. Based on this classification, we found that one of the stars, HD 331986, is a disc star exhibiting kinematic properties indicative of a transition between the thin and thick disc stellar populations (see also \citealt{matsunaga2022}) In contrast, the other two stars display dynamical properties characteristic of the halo (see Fig.~\href{https://zenodo.org/records/14610541}{D.1}). Atmospheric parameters and abundances are listed in Table \ref{table:abundances}. Note that v$_\gamma$ radial velocities are based on Gaia measurements. The v$_\gamma$
for HD331986 is based on the fit of the radial velocity time series, while for the
other two RRLs, it is the simple average of the radial velocity measurements \citep{clementini2023}.

The EMP RRLs display a consistent pattern in their $\alpha$-element abundances, with Mg, Ca, and Ti levels elevated to those characteristic of older stellar populations in our Galaxy. Notably, the [Ca/Fe] ratios are somewhat lower than the Mg and Ti values. This discrepancy arises because the only available Ca I line in our spectra is the 4226 \AA \, line, which is known to provide significantly underestimated Ca abundances in giants with equivalent widths (EW) greater than 60 m\AA. This was initially highlighted by \cite{mashonkina2007} and \cite{spite2012} and was extensively discussed by \cite{sitnova2019}. The [Al/Fe] ratio is widely recognised as one of the most reliable indicators for distinguishing between accreted and in-situ components of our Galaxy (see e.g. \citealt{hawkins2015}, \citealt{das2020}, \citealt{horta2021}, \citeyear{horta2023}). In our analysis, we found that the two metal-poor, halo-like RRLs exhibit [Al/Fe] ratios that are only slightly below the solar values. Notably, the presence of the Al I line at 3944 Å is detected in both stars, as shown in the middle and bottom panel in Figure \ref{fig:synth} for DO Hya and BPS CS 30317-056, respectively. This finding contrasts sharply with the star HD 331986, for which we could only establish a conservative upper limit of [Al/Fe] < 0 (upper panel of Figure \ref{fig:synth}).
To get further insights on the nature of our three EMP RRLs we compare our abundances for [Mg/Mn] vs [Al/Fe], as done in the literature for the identification of special structures in the Galactic disc, such as the ancient proto-building block $Loki$ by \cite{sestito2024} or Icarus \citep{refiorentin2024}. We also refer the reader to recent works by for example \citet{bellazzini2024}, \citet{sestito2024}, \citet{nepal2024}, and \citet[and references therein]{zhang2024}.

Our analysis is based on optical spectroscopy with real-time NLTE spectral synthesis calculations. Therefore, a suitable choice of large-scale comparison sample would be the similarly observed/analysed GALAH survey \citep{2015desilva}, Data Release 4 (DR4, \citealt{buder2024}).

We selected giant stars with the following quality flags:
\begin{itemize}
    \item \texttt{snr\_px\_ccd3} $>$ 70 (SNR per pixel at 6500  \AA $> $ 70) 
    \item \texttt{flag\_sp} = 0 (reliable spectroscopic analysis)
    \item \texttt{flag\_X\_fe} = 0 for Fe, Mg, Mn, and Al (reliable abundances)
    \item 3500 K $< T_{\rm eff} <$ 5500 K 
    \item $\ log g$ < 3.6
\end{itemize}

\begin{figure*}
    \centering
    \includegraphics[width=0.9\textwidth]{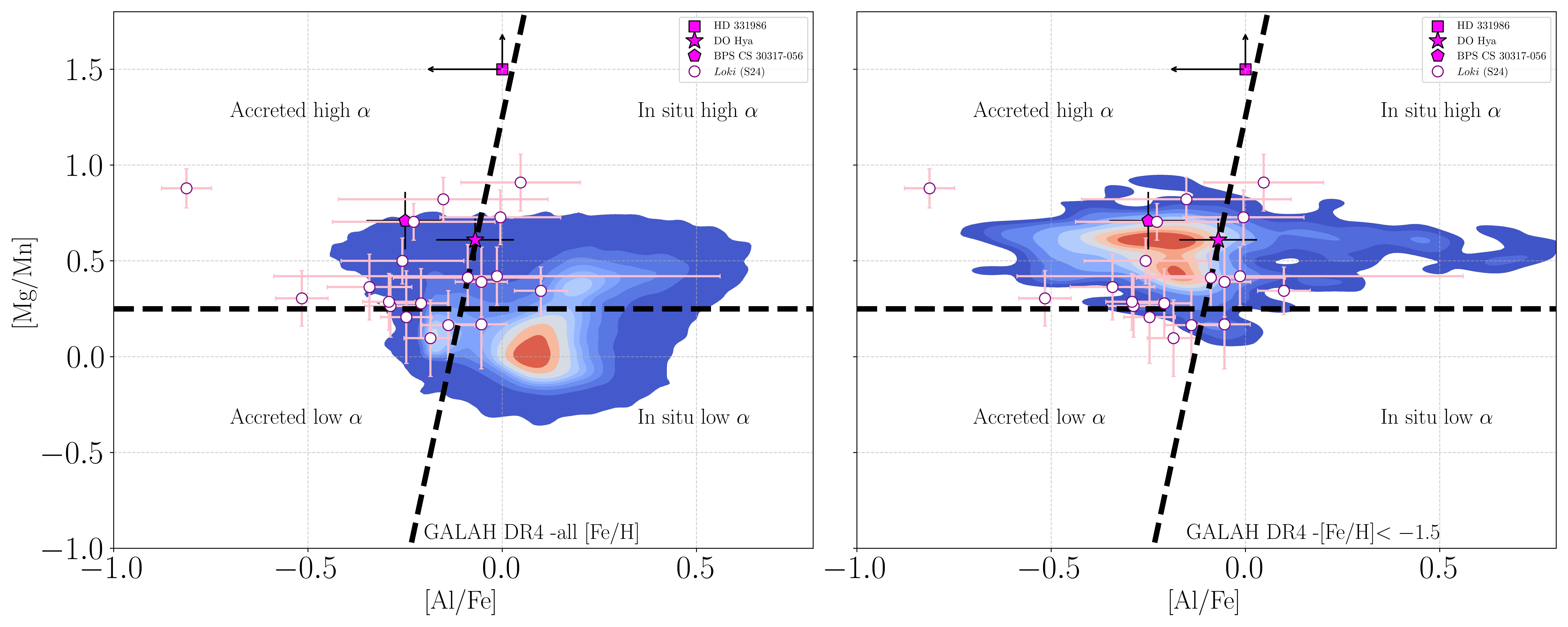}
    \caption{[Mg/Mn] ratios as a function of [Al/Fe] for our three RRLs (filled markers) along with literature estimates, including GALAH DR4 \citep[][density plots in this figure]{buder2024} and $Loki$ (empty circles) by \cite{sestito2024}.}
    \label{fig:MgMn_AlFe}
\end{figure*}

In the left-hand panel of Figure \ref{fig:MgMn_AlFe}, we display the relative abundances of [Mg/Mn] vs [Al/Fe] for all the selected stars from GALAH DR4, irrespective of their iron content ([Fe/H]), alongside our sample of EMP RRLs and the $Loki$ stars introduced by \cite{sestito2024}. These $Loki$ stars have not been adjusted to match the APOGEE scale, as done in the original paper. We have included dividing lines that help classify stars as either in situ or accreted, based on their [Al/Fe] and [Mg/Mn] abundance ratios, as proposed by for example \cite{sestito2024}, among the others. Most stars are situated within the low-$\alpha$ in situ stellar population (lower right-hand quadrant). However, there are notable concentrations in high-$\alpha$ in situ (upper right-hand quadrant) and low-$\alpha$ accreted regions (lower left-hand quadrant, but see discussion in \citealt{buder2024}).

When we shift our focus to the right-hand panel of Figure \ref{fig:MgMn_AlFe}, specifically examining stars with low metallicity ([Fe/H] $\lesssim -$ 1.5), these stars predominantly transition to the accreted high-$\alpha$ region. This observation indicates that the use of aluminium-based diagnostics is quite effective at identifying stellar origins at intermediate metallicities ([Fe/H] $\gtrsim -$1.5). However, their diagnostic power seems to diminish at even lower metallicities, as also highlighted by \cite{sestito2024}. Thus, our findings suggest that chemical planes involving combinations of aluminium, magnesium, and manganese do not prove useful for distinguishing accreted structures when dealing with very low metallicities. Consequently, the origin of our observed stars remains enigmatic, as no alternative chemical abundance plane appears viable for differentiating between in situ and accreted stars at these low metallicities. This limitation presents significant challenges for employing chemical tagging techniques on very metal-poor stars, which are vital for understanding the initial phases of galaxy formation and evolution.

\begin{figure}
    \centering
    \includegraphics[width=0.4\textwidth]{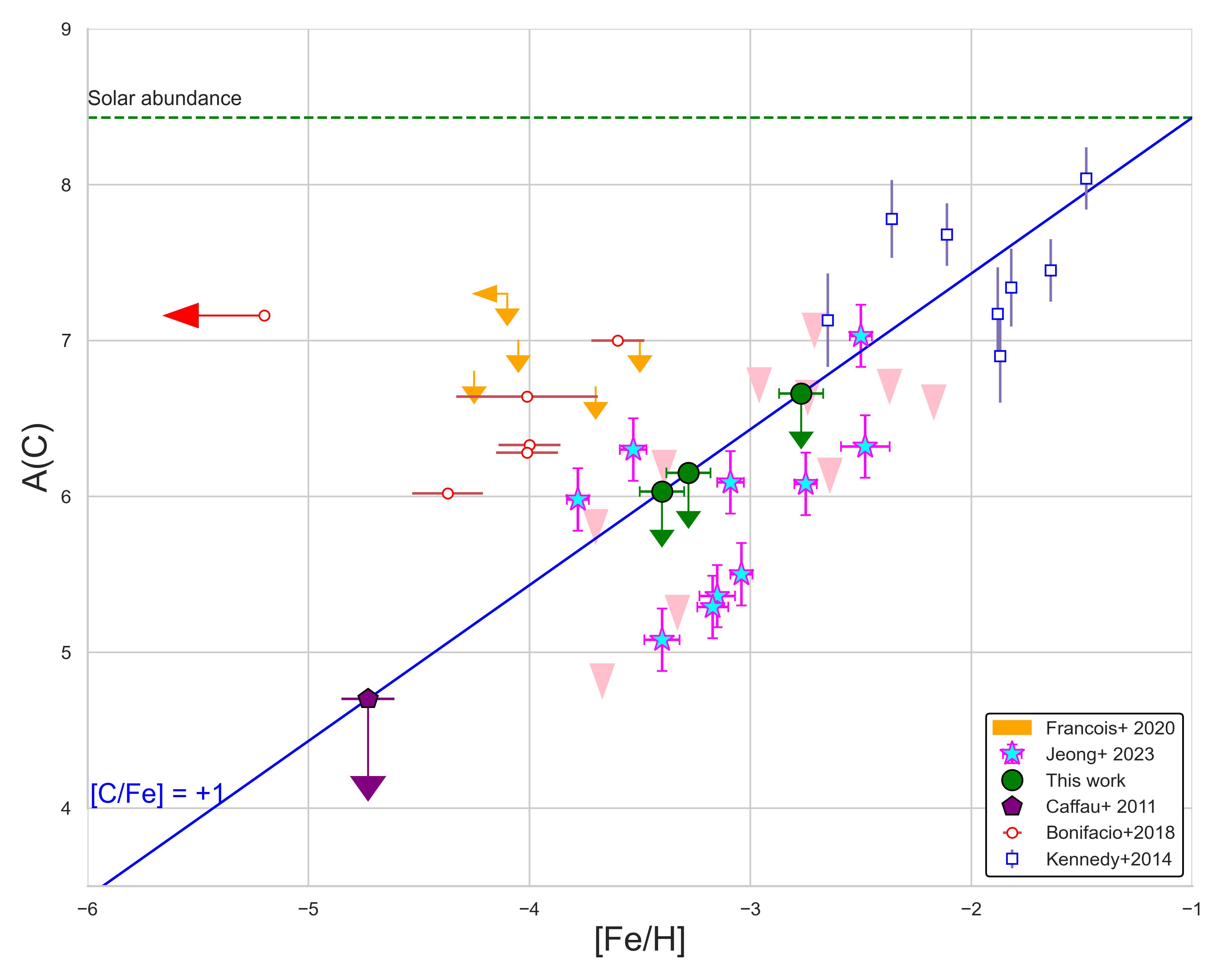}
    \includegraphics[width=0.4\textwidth]{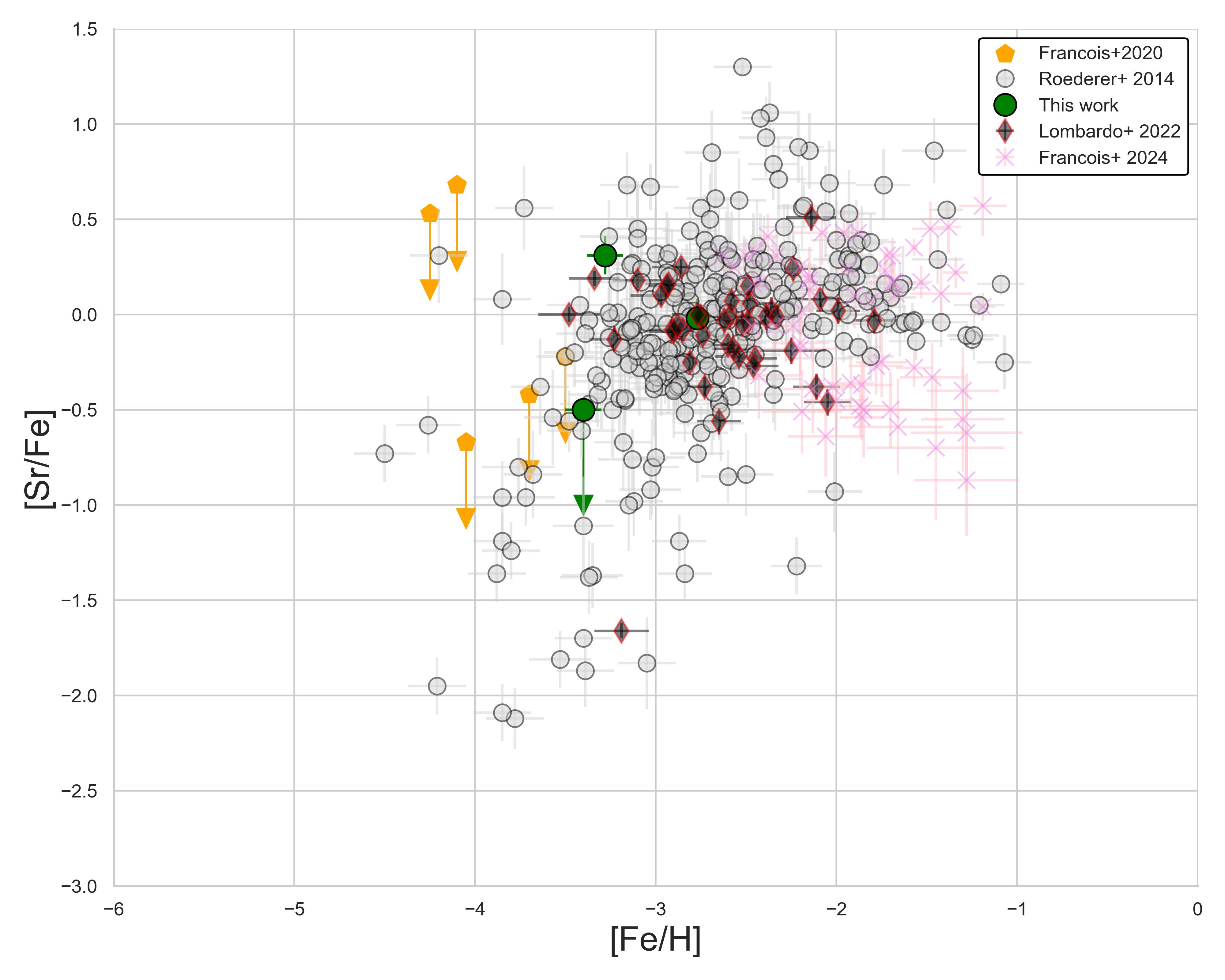}
    \caption{Carbon and Sr as a function of [Fe/H]. Upper panel: C abundances for our three EMP RRLs (green points) as a function of  [Fe/H]. For comparison, we include also stars from \cite{jeong2023, francois2020,bonifacio2018,caffau2011,kennedy2014}. 
    All abundances are in LTE; the blue solid line marks [C/Fe] = +1, see e.g.  \cite{beers2005}.
    Lower panel: [Sr/Fe] ratios vs. [Fe/H] for our RRLs along with literature samples \citep{roederer2014,francois2020,francois2024,lombardo2022}. We provide NLTE abundances as done by \cite{francois2024}, while LTE values were given by \cite{roederer2014, francois2020, lombardo2022}.}
    \label{fig:carbon}
\end{figure}

We were able to derive abundances for only one post-Fe peak element,
strontium (Sr, Z = 38). In the Solar system, Sr nucleosynthesis is mainly attributed to the weak and main components of the slow (s) n-capture process \citep{karakas2014}. However, at the metallicity of our stars, the r(apid)-process is very likely the dominant production source (see the recent review by \citealt{arcones2023} and references therein). HD 331986 is characterised by low [Sr/Fe], in contrast to the solar abundance pattern shown by BPS CS 30317-056 and the even super-solar levels inferred in DO Hya. The evolution of $n-$capture elements (e.g. Sr and Ba) at low metallicity is a debated topic, especially considering the significant spread observed at metallicities $\lesssim -2.5$ dex (see the review by \citealt{cowan2021}). This spread is illustrated in the lower panel of Figure \ref{fig:carbon}, where our EMP RRLs are compared to the [Sr/Fe] ratios as a function of metallicity for stars from \cite{roederer2014} and \cite{francois2020}.

In the upper panel, we compared our upper limits on carbon abundance with literature measurements. Despite the warm temperature and low metallicity of our stars hindering a robust derivation of carbon abundances via molecular features, we can exclude the possibility that these three stars are characterised by an overabundance in [C/Fe] ratios, as reported for very metal-poor carbon-enriched stars by several studies in the literature (see, e.g. \citealt{placco2014}, \citealt{bonifacio2018}, \citealt{arentsen2022}). The three EMP RRLs align with most of the RRLs referenced by \cite{kennedy2014}, but two of their seven RRLs show a carbon overabundance.

\subsection{Evolutionary timescales of EMP RR Lyrae stars}\label{sec:evolutionary}
\begin{figure*}
    \centering
    \includegraphics[width=0.49\textwidth]{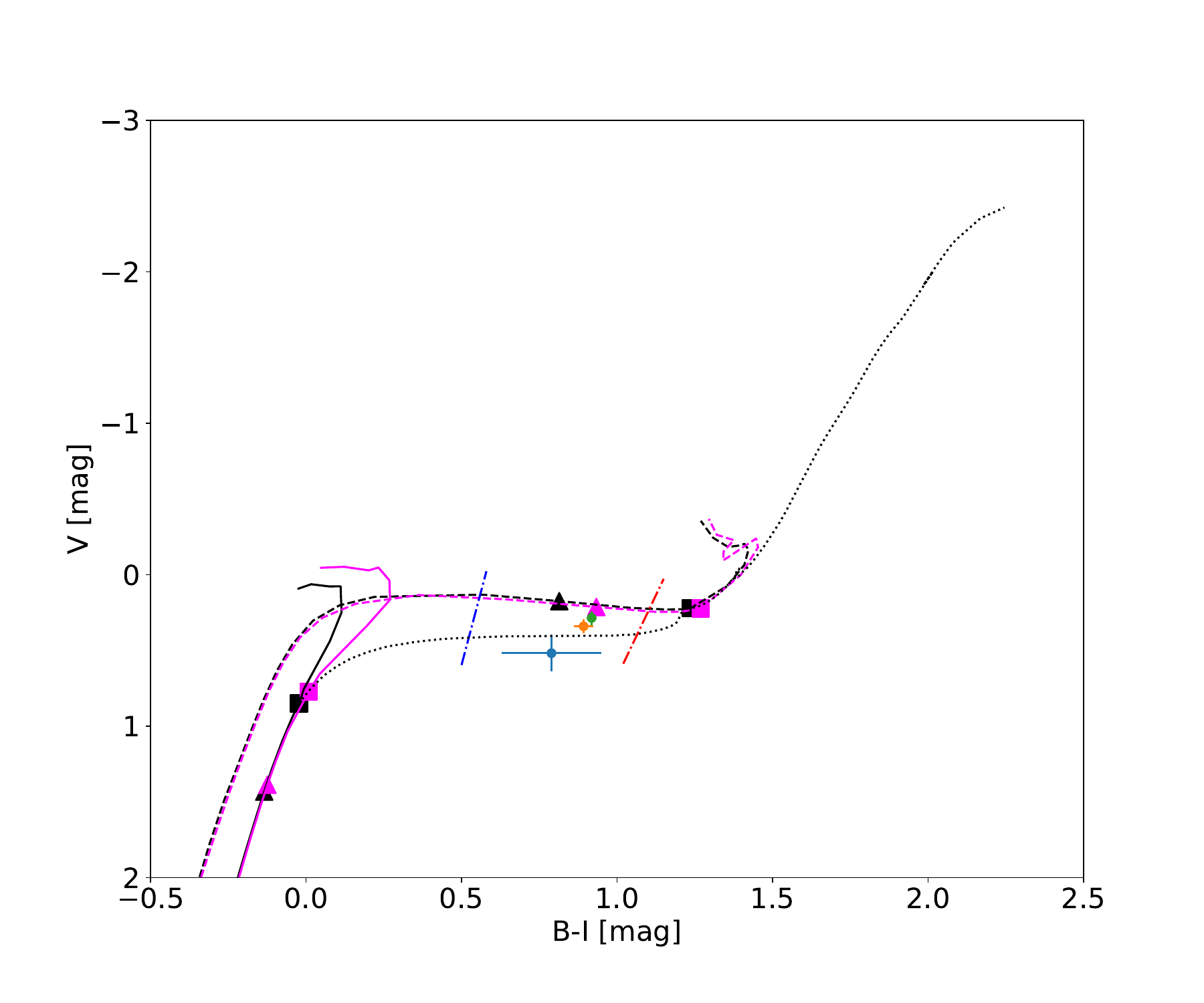}
    \includegraphics[width=0.49\textwidth]{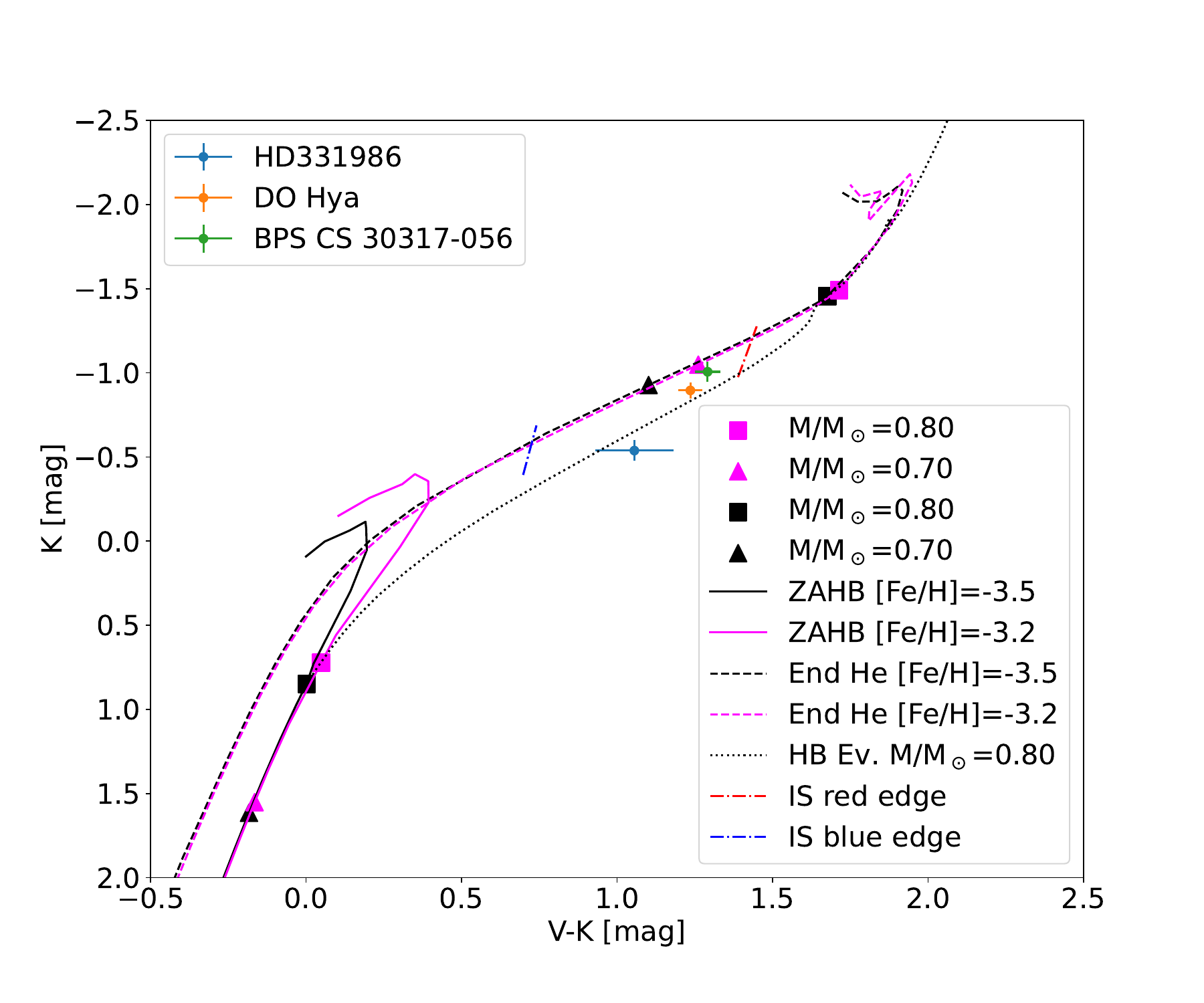}
    \caption{Left -- Optical $B$-$I$, $V$ CMD showing the location 
    of very metal-poor RRLs. The solid lines display predicted ZAHBs at different chemical compositions (see labelled values), while the 
    dashed lines the central helium exhaustion for the same chemical compositions
    \citep{pietrinferni2021}. 
    Squares and triangles mark different stellar masses along the ZAHB and 
    end-of-helium. The almost vertical lines display the blue (hot) and the red (cool)
    edge of the predicted RRL instability strip for Z=1$\times$10$^{-5}$ specifically 
    computed for this investigation (see text for more details).  
    Right -- Same as the left, but for the optical-NIR ($V$-$K$, $K$) CMD.}
    \label{fig:opt_nir_cmd}
\end{figure*}

Our investigation reveals that, despite intrinsic differences in heavy element composition, EMPs are as common among RRLs as they are in the
general field star population. Notably, evolutionary \citep{pietrinferni2021} and pulsation models (details are given in Appendix \ref{sec:appendix_models}) strongly support the existence of very metal-poor RRLs. The left panel of Fig.~\ref{fig:opt_nir_cmd} illustrates a comparison between very metal-poor evolutionary models, which assume an $\alpha$-enhanced chemical mixture with metal contents of Z = 1 $\times 10^{-5}$  and Z =  2 $\times 10^{-5}$, and the three EMP RRLs. To examine the positioning of the EMP RRLs, we plotted the zero-age horizontal branch (ZAHB, solid line) and the end of helium burning (dashed line). It is anticipated that low-mass, central helium-burning stars will spend the majority of their helium-burning phase between the ZAHB and the end-of-helium lines. Different symbols indicate the locations of two distinct stellar masses along the ZAHB and at the end of helium burning (as labelled). Furthermore, the blue and red vertical lines display the blue and red edges of the RRL instability strip (IS). Note that calculations provided by \citet{marconi2015} did not include the very metal-poor regime (Z$\ge$$10^{-4}$) and they were specifically computed for this investigation. Reddening, distances and mean magnitudes of the three EMP RRLs were adopted from different sources, depending on the star. For Figure \ref{fig:opt_nir_cmd} the optical mean magnitudes for HD 331986 are based on images collected with IAC80 telescope (see Appendix~\ref{sect:appendix_phot}), while NIR mean magnitudes are based on 2MASS measurements and light curve templates \citep{braga2019}. The distance is based on Gaia DR3 parallax, according to \cite{bailer-jones2021}. Moreover, we adopted a reddening estimate by Matsunaga (private communication) and the reddening law by \citet{wang2019}. 

For DO Hya and BPS CS 22881-039 the optical mean magnitudes come from ASAS-SN and Gaia $GB_pR_p$ transformed using \citet{pancino2022}, while NIR magnitudes from 2MASS. The distance was estimated using the PL in the WISE $W1$ pass--band \citet{mullen2023}, the reddening from \citet{schlafly11} and the reddening law from \citet{cardelli89}. The left panel of the same figure shows the comparison between theory and observations but in the optical-NIR $V$-$K$, $K$ colour-magnitude diagrams (CMD). The $V$-$K$ is more sensitive to temperature variations and the $K$ is one order of magnitude less affected by uncertainties in reddening correction. 

The data presented in Fig.~\ref{fig:opt_nir_cmd} demonstrate a remarkable agreement between theoretical predictions and observations. The three EMP RRLs appear to behave as canonical HB stars during their evolution off the ZAHB, with the ZAHBs in this extremely metal-poor range reaching effective temperatures that are generally hotter than the RRL IS (see Appendix \ref{sec:appendix_models} for further details). The main finding from comparing the evolutionary lifetimes of EMP and metal-poor stars within the instability strip is that the difference is only a factor of two to three. This suggests that we should be able to detect more EMP RRLs as larger catalogues (see e.g. \citealt{medina2024}) and samples of spectra become available, such as those from upcoming large-scale surveys such as 4MOST \citep{dejong2019} and WEAVE \citep{jin2024}.

\section{Concluding remarks}

This study presents a detailed spectroscopic analysis of three EMP RRLs, pushing the boundaries of known metallicity for this old stellar tracer.  Our high-resolution spectroscopic NLTE analysis yielded abundances for key elements (Fe, Al, Mg, Ca, Ti, Mn, and Sr).  Kinematic analysis revealed a diverse origin for these stars: HD 331986 exhibits disc kinematics, while DO Hya and BPS CS 30317-056 are consistent with halo populations, highlighting the ubiquity of such very metal-deficient variable stars across the Galactic components.  

Interestingly, the commonly used [Al/Fe] and [Mg/Mn] ratios, effective in distinguishing accreted and $in-situ$ stellar populations at higher metallicities, proved less effective for [Fe/H] $\lesssim -$ 1.5. This challenges the current understanding of Galactic chemical evolution at the lowest metallicities and underscores the need for more sophisticated chemical tagging techniques. The observed [Sr/Fe] ratios display considerable variation amongst the three stars, suggesting a complex nucleosynthesis history and reinforcing the limitations of using single elemental ratios for tracing stellar origins. The results also highlight the need for further investigations into the production of r-process elements at extremely low metallicities.

Our theoretical models strongly support the existence of EMP RRLs; further observational studies are required to identify more EMP RRLs and refine our understanding of Galactic chemical evolution and stellar nucleosynthesis at extremely low metallicities.

\section*{Data availability}

Appendices D-E are available on Zenodo (\href{https://zenodo.org/records/14610541}{at this link})

\begin{acknowledgements}
This research was supported by the Munich Institute for Astro-, Particle and BioPhysics (MIAPbP) which is funded by the Deutsche Forschungsgemeinschaft (DFG, German Research Foundation) under Germany´s Excellence Strategy – EXC-2094 – 390783311.
M.M. and M.S.B. acknowledge support from the Agencia Estatal de Investigaci\'on del Ministerio de Ciencia e Innovaci\'on (MCIN/AEI) under the grant "RR Lyrae stars, a lighthouse to distant galaxies and early galaxy evolution" and the European Regional Development Fund (ERDF) with reference PID2021-127042OB-I00.
This article is based on observations made with the IAC80 operated on the island of Tenerife by the Instituto de Astrofísica de Canarias in the Spanish Observatorio del Teide.
SB acknowledges support from the Australian Research Council under grant number DE240100150. GI acknowledges support for this project from “La Caixa” Foundation (ID 100010434) under grant agreement “LCF/BQ/PI24/12040020”. AB acknowledges support for this project from the European Union's Horizon 2020 research and innovation program under grant agreement No 865932-ERC-SNeX. C.E.M.-V. is supported by the international Gemini Observatory, a program of NSF NOIRLab, which is managed by the Association of Universities for Research in Astronomy (AURA) under a cooperative agreement with the U.S. National Science Foundation, on behalf of the Gemini partnership of Argentina, Brazil, Canada, Chile, the Republic of Korea, and the United States of America.
Part of this work is based on archival data, software or online services provided by the Space Science Data Center - ASI. In particular, the GaiaPortal access tool was used for this research (\url{http://gaiaportal.ssdc.asi.it}). We would like to express our gratitude to the reviewer for their thorough examination of the manuscript and for their insightful comments and suggestions, which have enhanced the quality of our work.
\end{acknowledgements}

%
%
\bibliographystyle{aa} 
\bibliography{emp-rrl} 
%

%
%

\begin{appendix} 
\onecolumn

\begin{landscape}

\section{Atmospheric parameters and abundances for our three EMP stars}

In Table \ref{table:abundances} we provide atmospheric parameters and elemental abundances for the stars HD 331986, DO Hya, and BPS CS 30317-056. 
It includes $T_{\rm eff}$, logg, and $V_{\rm mic}$, along with  [Fe/H], [Al/Fe], [Mg/Fe], [Ca/Fe], [Ti/Fe], [Mn/Fe], and [Sr/Fe], all derived in NLTE. 
Additionally, the table contains data on RV, pulsation periods, and classification types of each star, indicating their variability class with HD 331986 
categorised as type RRc, while DO Hya and BPS CS 30317-056 are classified as RRab. 

\begin{table*}[h!]
\caption{Atmospheric parameters, abundances, radial velocity, pulsation period and class of HD 331986, DO Hya and BPS CS 30317-056. All abundances are in NLTE with model atoms retrieved from the following sources:  iron \citep{bergemann2012a, semenova2020}, calcium \citep{mashonkina2017, semenova2020}, titanium \citep{bergemann2011}, manganese \citep{bergemann2019}, aluminium \citep{ezzeddine2018}, magnesium \citep{bergemann2017}, and strontium come from \citep{bergemann2012b, gerber2023}. The line list is available upon request.}
\setlength{\tabcolsep}{2.4pt}
\begin{tabular}{lccccccccccccl} 
\hline
\hline
star & Teff & logg & $V_{\rm mic}$& [Fe/H] & [Al/Fe] & [Mg/Fe] & [Ca/Fe] & [Ti/Fe] & [Mn/Fe] & [Sr/Fe] & RV & Period & class \\
     & (K)  & (dex) & (km s$^{-1}$) &  &  & &  &  &  &  & (km s$^{-1}$) & (days) & \\
\hline
& & & & & & & & & & & & \\
HD331986 & 6450$\pm$80 & 2.90$\pm$0.10 & 3.5$\pm$0.5&  $-3.40\pm0.05$ & $< 0$ &0.50$\pm$0.05 & 0.21$\pm$0.10 & ---- & $<-1.00$ & $< -$0.5 & $-97.2$$\pm$$2.9$ & 0.3711815 & {\tiny RRc} \\
DO Hya & 5835$\pm$80 & 2.50$\pm$0.10 & 3.1$\pm$0.5&  $-3.28\pm0.02$ & $-0.07\pm0.10$  & 0.55$\pm$0.03 & 0.22$\pm$0.10 & 0.61$\pm$0.02 &  $-$0.06$\pm$0.10 & 0.31$\pm$0.10 & $272.5$$\pm$$6.8$ & 0.7133254 & {\tiny RRab}\\
{\tiny BPS} & 6025$\pm$80 & 2.50$\pm$0.15 & 3.0$\pm$0.5 & $-2.77\pm$0.05 & $-0.25\pm$0.10 & 0.34$\pm$0.05 & 0.24$\pm$0.10 & 0.56$\pm$0.05 & $-0.37 \pm 0.12$ & $-0.02\pm$0.08 & $-58.9$$\pm$$4.6$ & 0.7484904 & {\tiny RRab}\\
{\tiny CS 30317-056} & & & & & & & & & & & & \\
\hline
\hline
\end{tabular}
\label{table:abundances}
\end{table*}
\end{landscape}

\section{Photometric properties}\label{sect:appendix_phot}

\begin{figure}
    \centering
    \includegraphics[width=0.49\textwidth]{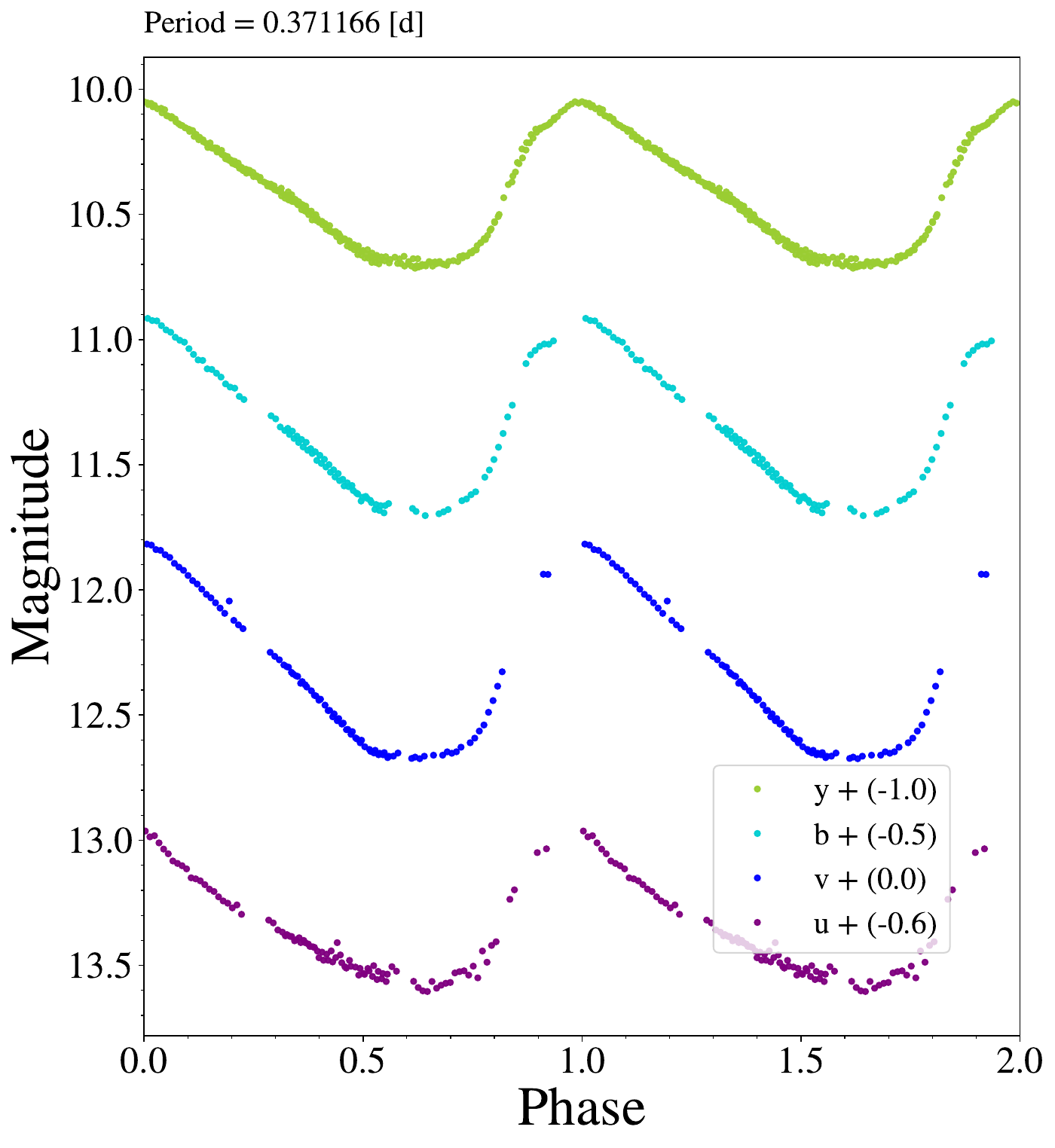}
    \caption{Light curves of HD 331986 in Str\"omgren \textit{ybvu} filters. Magnitudes other than \textit{b} are artificially shifted by the number indicated in the legend for the sake of a better representation in a single plot.}\label{fig:lightcurves}
\end{figure}

We performed optical photometry for HD 331986 using the IAC80 telescope located at the Teide Observatory in Tenerife (Spain). We used two different photometric systems (i.e. wide-band Johnson-Cousins UBVRI and mid-band Strömgren uvby) to constrain the phase of the spectroscopic observations and determine a first estimate of different stellar parameters from the pulsation properties. We obtained a total of 600 epochs for UBVRI and 100 epochs for Str\"omgren, observing mainly during the summer of 2021. We carried out data reduction, aperture photometry and calibration of the 
scientific frames using a Python custom-based pipeline, astrometry.net \citep{lang10}, {\sc SCAMP} \citep{bertin06}, {\sc SExtractor} \citep{bertin96}, and {\tt gnuastro} \citep{akhlaghi18}. We standardised the photometry using the same pipeline, aligning it with the Landolt system for the Johnson-Cousins filters \citep{stetson00} and the Hauck \& Mermilliod system for the Str\"omgren filters \citep{hauck98}. The precision achieved in the photometry was between 0.01 and 0.03 magnitudes for the UBVRI passbands and between 0.02 and 0.10 magnitudes for the Str\"omgren passbands. This precision decreases toward the bluer filters due to the lower quantum efficiency of the CCD. The resulting light curve is presented in Figure \ref{fig:lightcurves}. 

\begin{figure}
    \centering
    \includegraphics[width=0.49\textwidth]{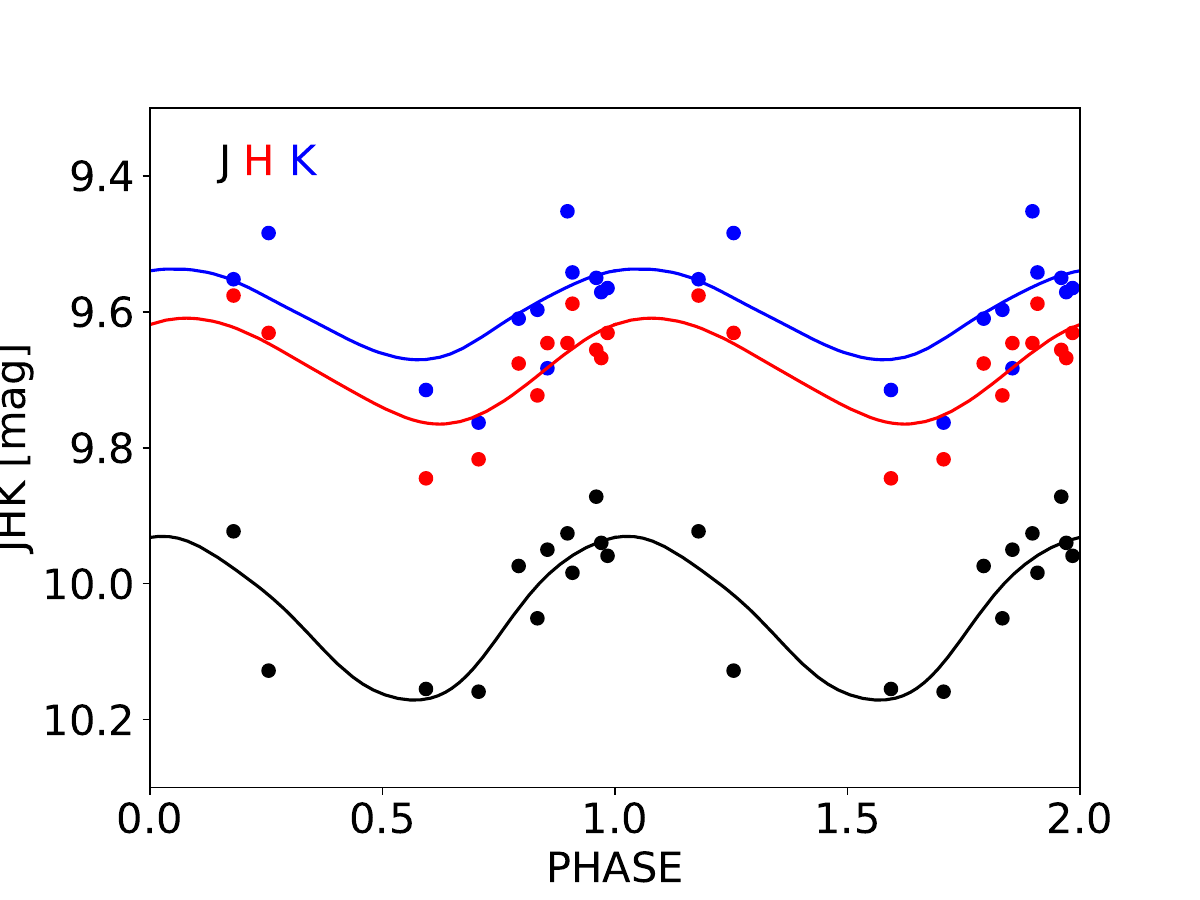}
     \caption{Black, red and blue circles represent the $J-, H-$ and $K-$ band data for HD 331986 collected with NIC at Nayuta Telescope in November 2021. Solid lines of the same colour represent the light curve templates by \cite{braga2019}, adopted to fit the empirical data.}
     \label{fig:NIR_lightcurve}
\end{figure}

We acquired near-infrared photometry for HD331986 using the Nishiharima Infrared Camera (NIC) on the 2.0 m Nayuta telescope at the Nishi-Harima Astronomical Observatory. The observations were made at 14 epochs in 2021 November. Reduction of the images followed a standard procedure for near-infrared images, including dark subtraction and flat-field correction. The aperture photometry was made also with SExtractor \citep{bertin1996} and the photometric zero points was determined by comparing the measurements of $\sim$20 stars in the $2^\prime .73 \times 2^\prime .73$ field-of-view with their JHKs magnitudes in the 2MASS catalogue \citep{skrutskie2006}. Figure~\ref{fig:NIR_lightcurve} shows the light curves.  The mean magnitudes were obtained with fitting light-curve templates \citep{braga2019}.

For the reddening estimate, we used an extinction of A$_V$=0.85 mag, derived by \citet{matsunaga21}. We adopted the reddening law from \citet{schlafly11}, obtaining the following corrections for each of the Strömgren passbands: 

$c_u$ = 1.61

$c_b$ = 1.25

$c_v$ = 1.41

$c_y$ = 1.00

Strömgren photometry can be used to define several colour indices that are sensitive to different stellar parameters, that is, temperature, surface gravity, and metallicity. These indices are defined as follow:

(b-y), for temperature

$c_1$ = (u-v)-(v-b), for surface gravity

$m_1$ = (v-b)-(b-y), for metallicity

As so, we computed both $c_1$ index and the reddening-sfree [$c_1$] (defined as [$c_1$] = $c_1$ - 0.19 (b-y)) for every epoch, before and after applying the reddening correction, obtaining the following mean values:

$c_1$ = 1.28$\pm$0.28

$c_1$ with de-reddened magnitudes $c_1$ = 1.25$\pm$0.28

[$c_1$] = 1.20$\pm$0.28

[$c_1$] with de-reddened magnitudes [$c_1$] = 1.22$\pm$0.28\\

\begin{figure*}
    \centering
    \includegraphics[width=0.9\textwidth]{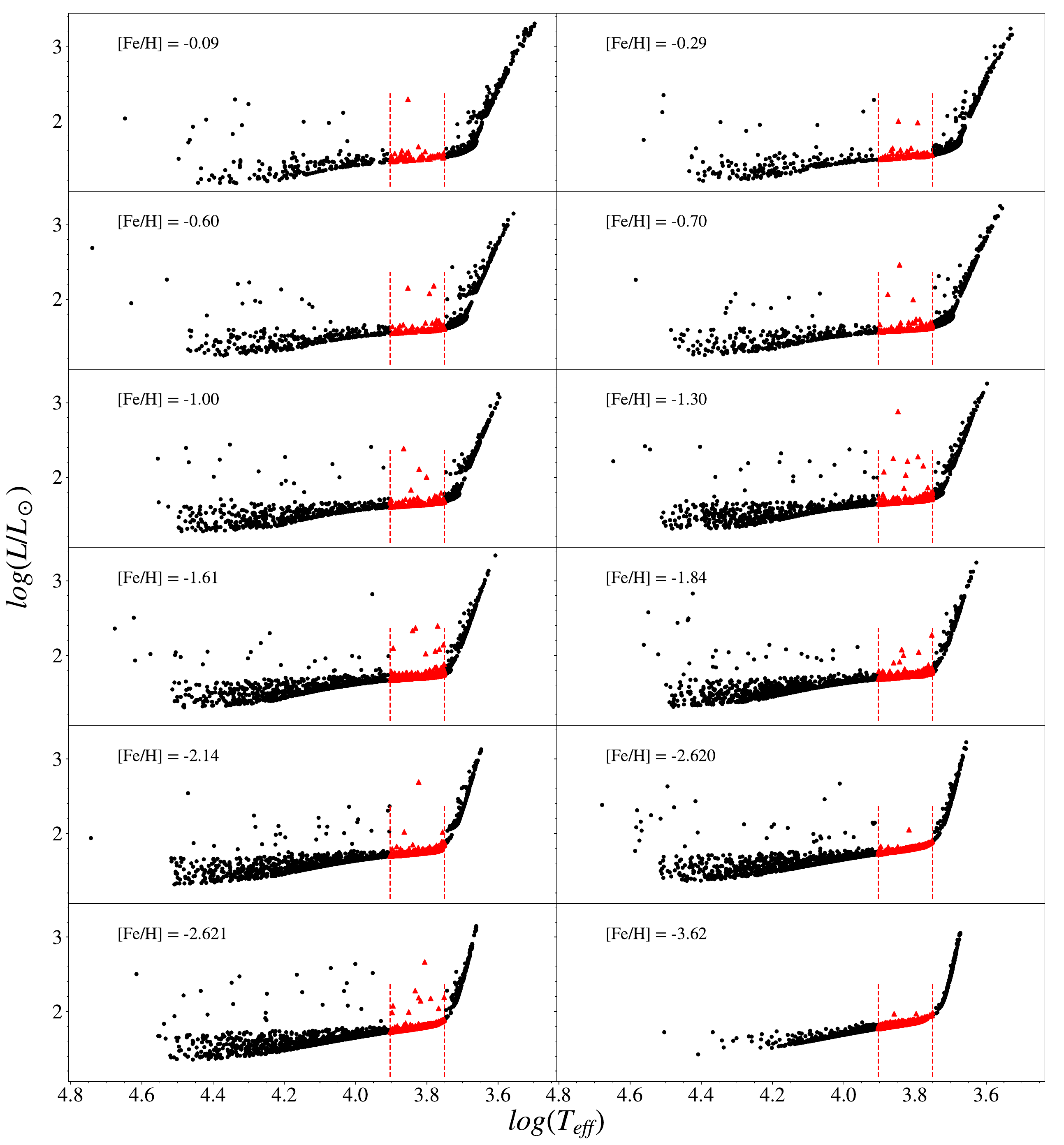}
    \caption{Hertzsprung Russell diagrams for the simulated HBs. Each panel shows a synthetic HB for different metal abundances (see labeled values). The dashed vertical lines display the blue (hot) and the  red (cool) edge of the RRL instability strip.
     }\label{fig:lum_te}
\end{figure*}

\begin{figure*}
    \centering
    \includegraphics[width=0.49\textwidth]{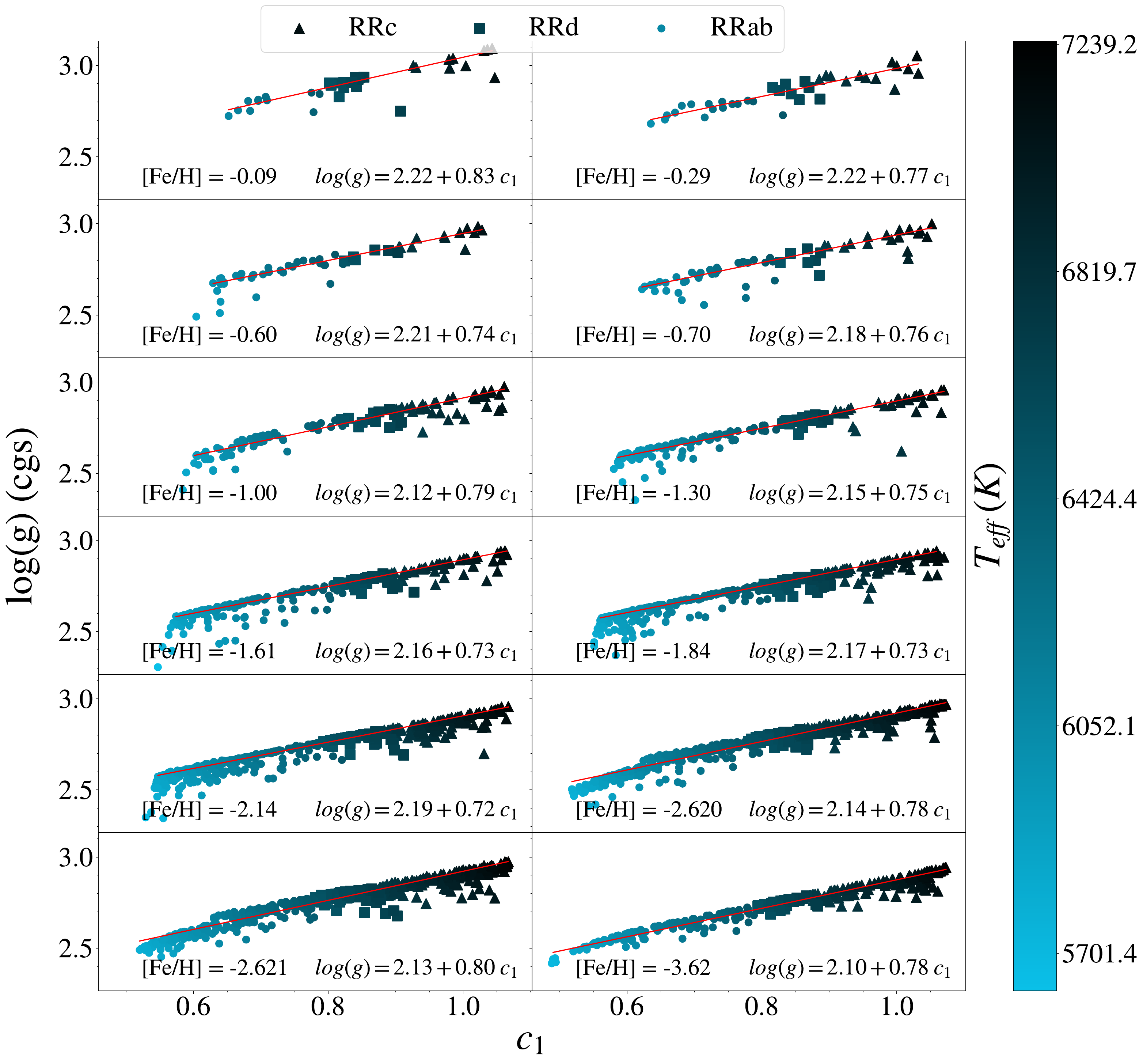}
    \includegraphics[width=0.49\textwidth]{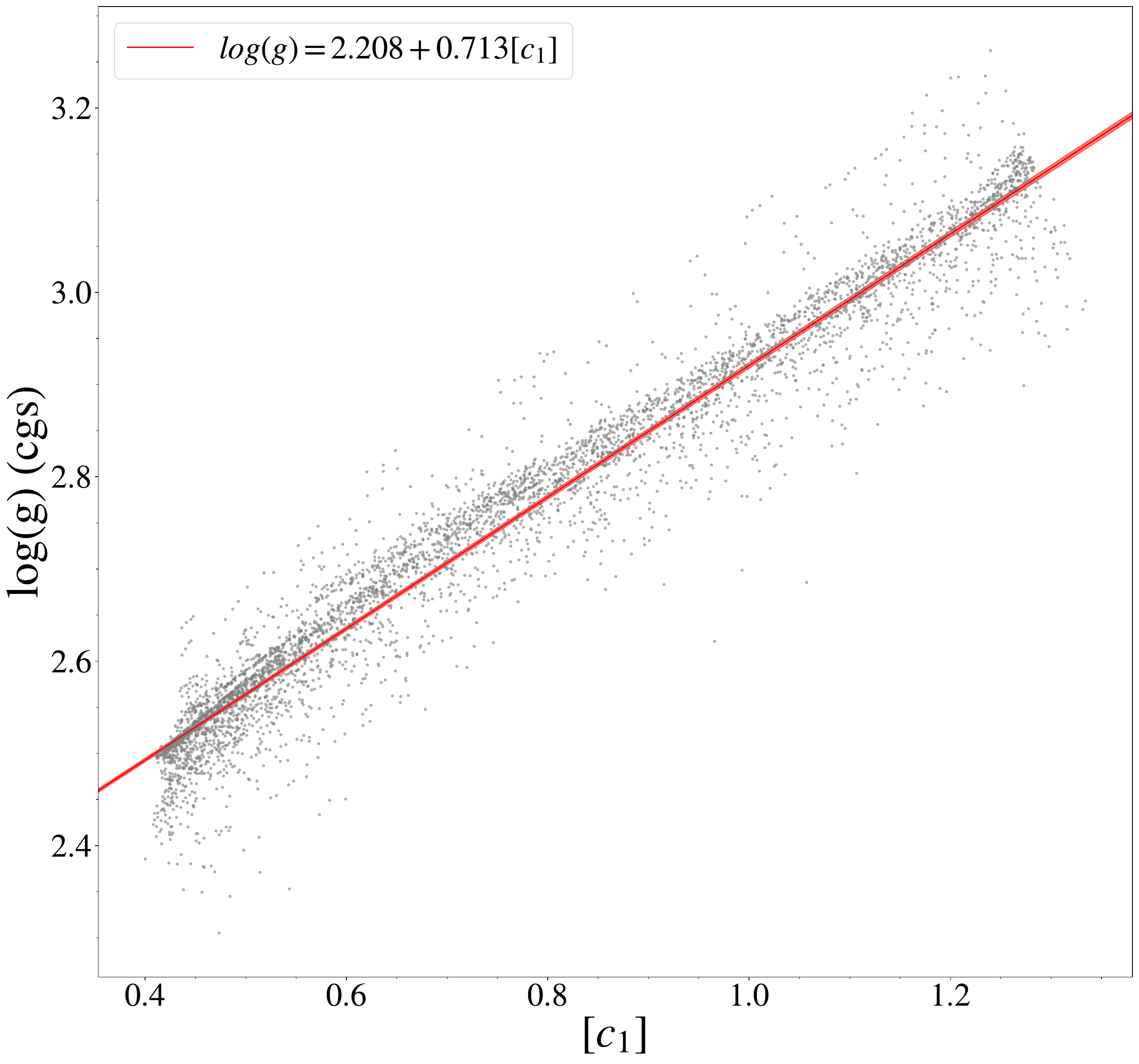}
    \caption{Left - $c_1$ coefficient as a function of surface gravity measured in cgs units. Each of the frames represents this relation for a different [Fe/H] in the range (-0.09, -3.62). Both the metallicity and the parameters of the fit are over-impressed in the bottom central area of each frame. Points are colour-coded according to the effective temperature of each star. Right - [$c_1$] coefficient as a function of surface gravity. In this plot, all the metallicities have been joined together to compute a relation for the whole set of parameters at once, given that dependency with [Fe/H] is negligible, as seen in the left panel.}\label{fig:c1}
\end{figure*}

To study the relation among $c_1$ index and surface gravity, we employed a set of synthetic CM diagrams covering a broad range in  metallicities ([Fe/H] = $-0.09, -3.62$), obtained using the BaSTI horizontal branch evolutionary models \citep{hidalgo18}. The synthetic HB models were adopted to investigate on a quantitative basis the possible dependence of the $c_1$-$\log g$ relation on the metal content for effective temperatures and surface gravities typical of RRLs. Figure \ref{fig:lum_te} shows the synthetic HB models for twelve different iron abundances and the red dots mark the predicted RRLs adopted to derive the c1-log g relation. We performed a linear fit over the entire data set, namely selecting stars located inside the RRL instability strip, i.e objects with an effective temperature ranging from $\sim$5600 K to $\sim$8000 K. Interestingly enough, we found no significant dependence on the iron abundance, as can be seen on the left panel of Figure \ref{fig:c1}. This shows the linear fit for each metallicity diagram together with a plot with the dependency of the slope and intercept with the metallicity. Taking this into account, we joined together all the CMDs to obtain a single and simpler relation between the surface gravity and the $c_1$/[$c_1$] indices, the second one being a reddening free index, as noted previously in the text. This is valid in the selected range of temperatures and has a linear dependency as follows:\\

$\log g$ (cgs) = m $c_1$ + n\\\\
m = 0.763$\pm$0.003\\
n = 2.131$\pm$0.002\\

$\log g$ (cgs) = m [$c_1$] + n\\\\
m = 0.713$\pm$0.002\\
n = 2.208$\pm$0.002\\

As a final step, surface gravity was computed from this relation and the $c_1$/[$c_1$] index that arose from photometry at the same phase where the spectra were gathered. There is excellent agreement between our photometric and spectroscopic surface gravity $\log g$ (difference less than 0.1 dex).\\

\section{Theoretical models}\label{sec:appendix_models}

\begin{figure*}
    \centering
    \includegraphics[width=0.9\textwidth]{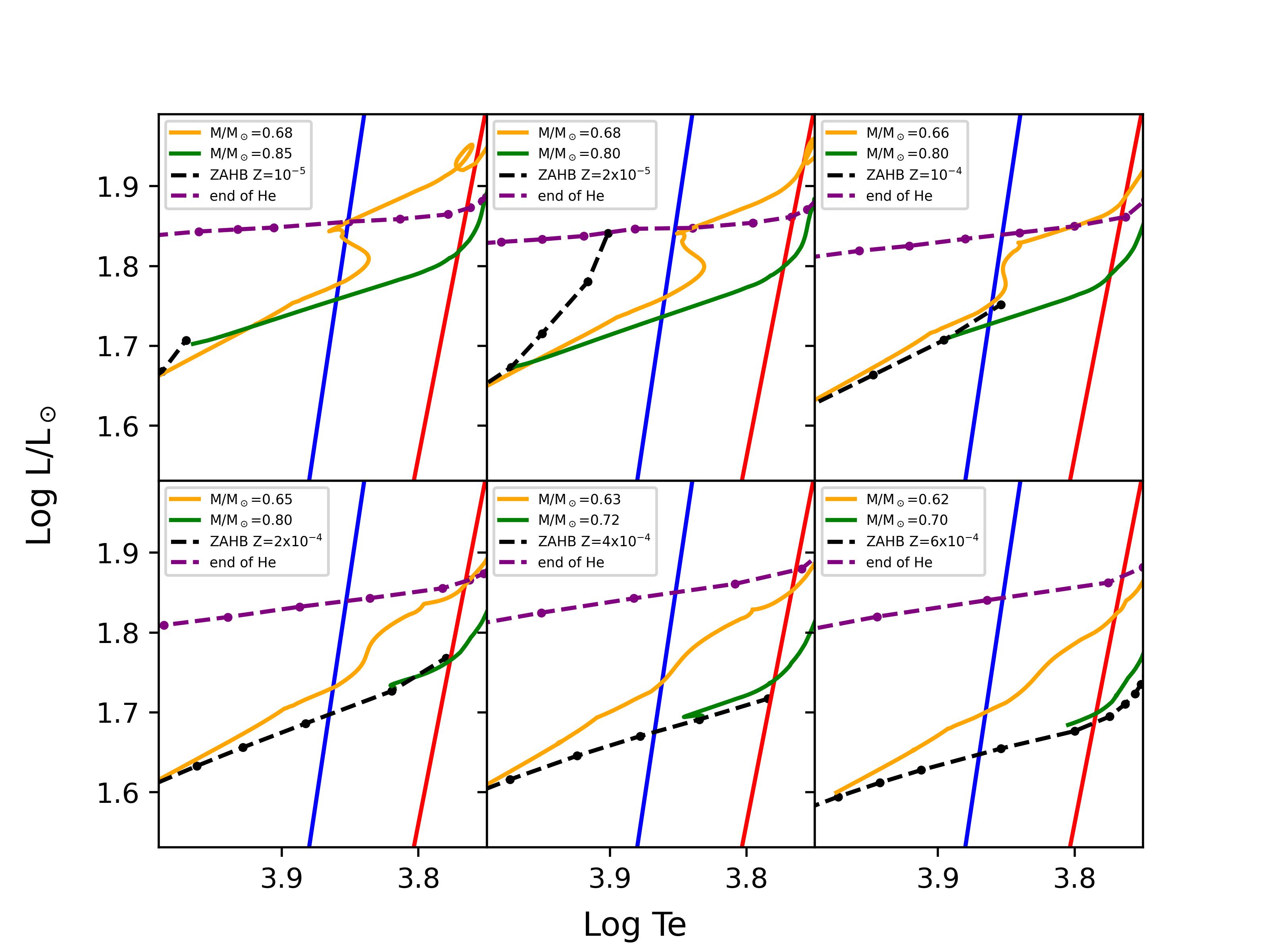}
    \caption{Hertzsprung-Russell diagram for low-mass stars
     during central helium-burning phases. The different panels display
     predictions for different metal-poor and extremely metal-poor 
     chemical compositions (see labelled values). The dashed lines show the 
     ZAHB (black) and the 
     end-of-central-helium burning (endHB, purple). Coloured lines are selected 
     HB evolutionary models showing the typical RRL crossing of the IS.  
     The green line shows the evolutionary path of a star close 
     to the ZAHB while the orange line the evolutionary path of a low-mass 
     evolved RRL. Masses that produced RRLs are 
    highlighted in the legend. The almost vertical blue and red 
    lines display the predicted RRL instability strip from \citet{marconi2015} 
    updated with extremely metal-poor pulsation models at Z=0.00001.}\label{fig:hrd_theo_meta}
\end{figure*}

To provide a more comprehensive theoretical framework for low-mass central 
helium-burning stars, Figure~\ref{fig:hrd_theo_meta}
shows in the Hertzsprung-Russell diagram predicted ZAHBs (black dashed lines), 
end-of-central-helium burning (purple dashed lines) and representative HB 
evolutionary models (solid lines) for a wide range of metal-poor and 
extremely metal-poor $\alpha$-enhanced chemical compositions (see labelled 
values). Note that the current ZAHBs have been constructed by assuming a progenitor age of 13.5~Gyr at the tip of the red giant branch. The location on the ZAHB and on the end-of-helium burning of this progenitor mass are marked with filled circles. 

The predicted instability strip for RRLs provided by \citet{marconi2015} only covers the metal-poor regime ([Fe/H]$\sim-$2.50, Z=1$\times$10$^{-4}$). Therefore, we performed detailed calculations, using non-linear pulsation models, 
including a time-dependent treatment of the convective transport, for 
an $\alpha$-enhanced chemical composition of [Fe/H]$\sim -$3.62 (Z=$1\times$10$^{-5}$). 
The new instability strip is plotted in the top left panel of Fig.~\ref{fig:hrd_theo_meta} 
and the edges are listed in Table \ref{table:is}, alongside their optical and NIR magnitudes.
The instability strip for [Fe/H]$\sim -$3.20 (Z=2$\times$10$^{-5}$) was 
linearly interpolated.

\begin{table*}
\caption{Edges of the instability strip.}
\setlength{\tabcolsep}{2.8pt}
\begin{small}
\begin{tabular}{lccccccccccc} 
\hline
\hline
Edge  &  M/Mo   &  logL/Lo   &  Te    &     $M_V$   &  $B-V$   &    $U-B$  &  $V-I$  &    $V-R$  &   $R-I $ &  $V-J$  &  $V-K$  \\  
  \hline
& & & & & \\
FOBE  & 0.85  &  1.78 &   7250 &   0.305 &  0.20292 &  0.055  & 0.318 &  0.140 &  0.177 &  0.530  & 0.698 \\
FBE   & 0.85  &  1.78 &   7050 &   0.317 &  0.23346 &  0.030  & 0.367 &  0.164 &  0.202 &  0.604  & 0.802 \\
FORE  & 0.85  &  1.78 &   6450 &   0.369 &  0.33261 & -0.045  & 0.519 &  0.238 &  0.280 &  0.846  & 1.140 \\
FRE   & 0.85  &  1.78 &   6050 &   0.415 &  0.41022 & -0.081  & 0.624 &  0.290 &  0.333 &  1.021  & 1.389 \\
FOBE  & 0.85  &  1.88 &   7150 &   0.054 &  0.21277 &  0.054  & 0.337 &  0.149 &  0.187 &  0.561  & 0.741 \\
FBE   & 0.85  &  1.88 &   6950 &   0.067 &  0.24373 &  0.029  & 0.387 &  0.173 &  0.213 &  0.637  & 0.847 \\
FORE  & 0.85  &  1.88 &   6550 &   0.103 &  0.31103 & -0.021  & 0.489 &  0.223 &  0.265 &  0.799  & 1.074 \\
FRE   & 0.85  &  1.88 &   5950 &   0.173 &  0.42928 & -0.075  & 0.649 &  0.303 &  0.345 &  1.063  & 1.449 \\
\hline
\hline
\end{tabular}
\end{small}
\label{table:is}
\end{table*}

The current evolutionary and pulsation prescriptions indicate that for iron 
abundances more metal-poor [Fe/H]$\sim -$2.20 (Z=$2\times$10$^{-4}$), the ZAHBs (black dashed lines) based on old progenitors either do not cross or minimally cross the instability strip, that is the ZAHB attains colours that are systematically bluer (hotter) than the predicted RRL instability strip. The consequence is that in the very metal-poor and metal-poor regime ([Fe/H]$\leq- $2.2) RRLs can only be produced by stars during their off-ZAHB evolution.

The consequence of the systematic shift toward higher effective temperatures in the extremely metal-poor regime is that these stars are expected to spend, on average, two to three times less time inside the IS compared to their metal-poor and metal-intermediate counterparts.
Evolutionary prescriptions listed 
in Table~\ref{table:t_he_is} indicate that given an increase in iron 
abundance of $\sim$1.5~dex, from [Fe/H]=$-$3.6 to [Fe/H]=$-2.2$, the  
evolutionary time spent inside the IS ranges from $\sim$25--30~Myr 
in the very metal-poor regime to $\sim$70~Myr in more metal-rich 
stars. Note that in these
preliminary estimates we are assuming that the mass distribution 
inside the IS is linear over the entire range in effective temperatures covered by the IS. Moreover, we are also assuming that 
the star formation rate in moving from the very metal-poor 
stars to the more metal-rich ones is constant in time. 
Despite these assumptions, the statistics of EMP RRLs, which occur at a rate well below the 1\%, align with the occurrence rate of stars with metallicity [Fe/H] $\lesssim -$ 2.75 (see e.g. \cite{andrae2023}) and there is no significant tension as we observe consistent numbers.

\begin{table}
\caption{Evolutionary time spent inside the instability strip for different chemical compositions.}
\setlength{\tabcolsep}{2.8pt}
\begin{small}
\begin{tabular}{lcllrrr} 
\hline
\hline
  M            &   Y   &  Z   & [Fe/H] & t(IS$_{end}$) & t(IS$_{start}$) & (IS$_{cross}$)\\
  M$_{\odot}$  &       &      &        & Myr           & Myr             & Myr\\
  \hline
& & & & & \\
   0.68 & 0.24 &   0.00001 &   -3.62  & 77.5 &   69.2  &   8.3   \\
   0.85 & 0.24 &   0.00001 &   -3.62  & 62.6 &   44.5  &  18.0   \\
   0.68 & 0.24 &   0.00002 &   -3.20  & 78.6 &   68.7  &   9.9   \\
   0.80 & 0.24 &   0.00002 &   -3.20  & 65.9 &   47.8  &  18.1   \\
%
   0.66 & 0.24 &   0.0001  &   -2.50  & 84.6 &   78.2  &   6.4   \\
   0.80 & 0.24 &   0.0001  &   -2.50  & 67.9 &   43.3  &  24.7   \\
%
   0.65 & 0.25 &   0.0002  &   -2.20  & 87.2 &   78.3  &   8.9   \\
   0.80 & 0.25 &   0.0002  &   -2.20  & 61.9 &    1.0  &  60.9   \\
%
   0.63 & 0.25 &   0.0004  &   -1.90  & 92.5 &   87.5  &   5.0   \\
   0.72 & 0.25 &   0.0004  &   -1.90  & 70.4 &    1.0  &  69.4   \\
%
   0.62 & 0.25 &   0.0006  &   -1.71  & 93.4 &   82.7  &  10.7   \\
   0.70 & 0.25 &   0.0006  &   -1.71  & 63.7 &    1.0  &  62.7   \\
\hline
\hline
\end{tabular}
\end{small}
\label{table:t_he_is}
\end{table}

\end{appendix}

\end{document}